\documentclass[twocolumn,epjc3]{svjour3}
\smartqed  
\RequirePackage{graphicx}

\RequirePackage{amsfonts}
\RequirePackage{mathptmx}      
\RequirePackage{flushend}
\RequirePackage[numbers,sort&compress]{natbib}
\RequirePackage[colorlinks,citecolor=blue,urlcolor=blue,linkcolor=blue, pdfstartview={XYZ null null 1.00}]{hyperref}
\journalname{Eur. Phys. J. C}
\begin{document}

\title{Scattering of Dirac fermions by spherical massive bodies
}


\author{Ion I. Cot\u aescu\thanksref{e1,addr1}
        \and
           Ciprian A. Sporea\thanksref{e3,addr1}
}

\thankstext{e1}{e-mail: i.cotaescu@e-uvt.ro}

\thankstext{e3}{e-mail: ciprian.sporea@e-uvt.ro}

\institute{West University of Timi\c soara, V. P\^ arvan Ave. 4, RO-300223, Timisoara , Romania\label{addr1}}


\date{Received: date / Accepted: date}

\maketitle

\begin{abstract}
The asymptotic form of Dirac spinors in the field of a Schwarzschild black hole
is used for deriving analytically for the first time the phase shifts of the partial wave analysis of Dirac fermions scattered from massive spherical bodies, imagined as black holes surrounded by a surface producing total reflection. A simple model is analyzed by using graphical methods.
\end{abstract}

\section{Introduction}

In general relativity the study of the scattering of the scalar, electromagnetic or Dirac particles from Schwarzschild \cite{FHM,unruh,FHM1,ref1,ref2,S1,S2,bh1,S3}  black holes remains of actual interest.   An effective analyse of the scattering of the massive Dirac fermions from  Schwarzschild black holes was performed combining analytical and numerical methods  \cite{S1}-\cite{S3}.  Recently, starting with a set of asymptotic solutions \cite{CA},  we propsed  an analytic version of partial wave analysis that allowed us to write down closed formulas for the phase shifts giving the scattering amplitudes and cross sections of the fermions scattered from  Schwarzschild \cite{CCS}, Reissner-Nordstr\" om \cite{CCS1,CS1}, Bardeen \cite{CS2} and MOG \cite{CS3} black holes. Other studies that investigate fermion scattering by different types of spherically symmetric black holes can be found for example in Refs. \cite{Das, Jin, Doran, Jung, Gaina, Gaina1, ChaoLin, Rogatko, Liao, Ghosh}.

In the present paper we would like to extend this analytic study to the problem of the Dirac fermions scattered from a massive spherical body whose exterior radius is larger that the   Schwarzschild radius. We consider that this body is electrically neutral generating only a gravitational field with a Schwarzschild metric in its exterior while its surface is able to prevent the natural black hole absorption, reflecting  entirely the incident fermion beam.

Recently, in Ref. \cite{sdolan} the authors have studied the elastic scattering of a scalar field in the curved spacetime of a compact object. Moreover, studies of gravitational waves by compact objets also received little attention \cite{GWs1,GWs2,GWs3}.

Our method is based on the approximative solutions of the Dirac equation in the Schwarzschild geometry \cite{CA} that can be used at large distances from the black hole singularity \cite{CA,CCS}. This is not an impediment when the body is extended enough since then we can use these solutions for fixing different boundary conditions on the exterior surface. In general, the boundary conditions are  linear relations between these radial functions. Here we show that there exists a specific boundary condition determining the {\em total} reflection,  preventing absorption in any  partial waves where the natural black hole absorption might be possible \cite{CCS}.

Our principal goal here is to study the reflected electron beam by using the partial wave analysis for studying the cross sections and induced polarization. We use an analytical method based on the asymptotic approximation of the radial functions that allows us to derive for the first time closed formulas of the phase shifts and scattering amplitudes  in the case of the total reflection on a massive body. Thus we complete our analytical study devoted to the fermion scattering from black holes \cite{CCS,CCS1,CS1} obtaining new formulas which are in accordance with our previous results.

We must specify that our method based on the asymptotic approximation leads to results which are independent on the exterior radius of the massive body.  This creates an {\em ideal} image of the target which is seen as a point-like particle producing total reflection regardless its physical dimensions. This is somewhat  in accordance with the philosophy of the partial wave analysis which exploits mainly the asymptotic zone. Obviously, this ideal image can be corrected at any time by using numerical methods for which the analytical results we present here could be an useful guide.  Note that our preliminary numerical investigations show that these corrections  are very small remaining thus less relevant.

Under such circumstances, the reflecting surface cannot be characterized {\em a priori}  in terms of scattering parameters such that we must analyze different models by fixing the set the of arbitrary phases of the normalization factors and studying then the resulted scattering intensity and induced polarization for understanding the physical content of the model. Here we consider a simple model that can be compared with the model of the  bare black hole we studied earlier \cite{CCS}.

We start in the second section revisiting our approximative scattering solutions of the Dirac equation in Schwarzschild's geometry \cite{CA} which are used for developing the partial wave analysis briefly presented in the next section. In the fourth section the boundary conditions corresponding to the total reflection are proposed calculating the phase shifts for which we obtain simple closed formulas. The next section is devoted to the study of the models we propose here using graphical methods. In the last section we present our concluding remarks.  In what follows we use the notations of Refs. \cite{CCS,CCS1} and Planck's natural units with $G=c=\hbar=1$.

\section{Scattering  Dirac spinors in Schwarzschild's  geometry}

On Riemannian spacetimes the Dirac equation is defined in frames $\{x;e\}$  formed by a local chart of coordinates $x^{\mu}$, labeled by natural indices, $\alpha,..,\mu, \nu,...=0,1,2,3$, and an orthogonal  local frame and corresponding coframe  defined by the gauge  fields (or tetrads), $e_{\hat\alpha}$ and  respectively $\hat e^{\hat\alpha}$, labeled by the local indices $\hat\alpha,..,\hat\mu,...$ with the same range. In any local-Minkowskian manifold, $(M,g)$, having as flat model the Minkowski spacetime $(M_0,\eta)$ with the metric
\begin{equation}
\eta={\rm
diag}(1,-1,-1,-1),
\end{equation}
the gauge fields satisfy the  usual duality
conditions, $\hat e^{\hat\mu}_{\alpha}\,
e_{\hat\nu}^{\alpha}=\delta^{\hat\mu}_{\hat\nu},\,\, \hat
e^{\hat\mu}_{\alpha}\, e_{\hat\mu}^{\varphi}=\delta^{\varphi}_{\alpha}$
and the orthogonality relations, $e_{\hat\mu}\cdot
e_{\hat\nu}=\eta_{\hat\mu \hat\nu}\,,\, \hat e^{\hat\mu}\cdot \hat
e^{\hat\nu}=\eta^{\hat\mu \hat\nu}$. The gauge fields define the
1-forms $\omega^{\hat\mu}=\hat e^{\hat\mu}_{\nu}dx^{\nu}$ giving the
line element
$ds^2=\eta_{\hat\alpha\hat\varphi}\omega^{\hat\alpha}\omega^{\hat\varphi}=g_{\mu\nu}dx^{\mu}dx^{\nu}$.

The Dirac equation, $ (i\gamma^{\hat\alpha}D_{\hat\alpha} - m)\psi=0$
of a free spinor field $\psi$  of  mass $m$, may be written with our previous notations \cite{CCS,CCS1} in the frame $\{x;e\}$ defined by the Cartesian gauge,
\begin{eqnarray}
\omega^0&=&w(r)dt \,,\\
\omega^1&=&\frac{1}{w(r)}\sin\theta\cos\phi \,dr+ r\cos\theta\cos\phi \,d\theta\nonumber\\
&&- r \sin\theta\sin\phi \,d\phi\,, \\
\omega^2&=&\frac{1}{w(r)}\sin\theta\sin\phi \,dr+ r\cos\theta\sin\phi \,d\theta\nonumber\\
&&+ r \sin\theta\cos\phi \,d\phi\,, \\
\omega^3&=&\frac{1}{w(r)}\cos\theta \,dr- r \sin\theta \,d\theta\,,
\end{eqnarray}
in the gravitational  field of a  spherical body of mass $M$  with  the Schwarzschild line element
\begin{eqnarray}
ds^{2}&=&\eta_{\hat\alpha\hat\varphi}\omega^{\hat\alpha}\omega^{\hat\varphi}\nonumber\\
&=&w(r)^2dt^{2}-\frac{dr^{2}}{w(r)^2}- r^{2} (d\theta^{2}+\sin^{2}\theta~d\phi^{2})\,,\label{(le)}
\end{eqnarray}
defined on the radial domain $D_{r}= (r_{0}, \infty)$ where
\begin{equation}
w(r)=\left[1-\frac{r_0}{r}\right]^{\frac{1}{2}}\,,\quad r_{0}=2M\,.
\end{equation}
In what follows we study the scattering solutions of the Dirac equation in the asymptotic domain where $r\gg r_0$.

In the tertad-gauge  we consider here the spherical variables of the Dirac equation can be separated  just as in the case of the central problems in Minkowski  spacetime \cite{TH}, obtaining fundamental solutions of the form
\begin{eqnarray}
&&U_{E,\kappa,m_{j}}({x})=U_{E,\kappa,m_{j}}(t,r,\theta,\phi)=\frac{e^{-iEt}}{rw(r)^{\frac{1}{2}}}\nonumber\\
&&\times[f^{+}_{E,\kappa}(r)\Phi^{+}_{m_{j},\kappa}(\theta,\phi)
+f^{-}_{E,\kappa}(r)\Phi^{-}_{m_{j},\kappa}(\theta,\phi)]\,,\label{(u)}
\end{eqnarray}
expressed in terms of radial wave functions,  $f^{\pm}_{E,\kappa}$, and  usual four-component angular spinors $\Phi^{\pm}_{m_{j}, \kappa}$ \cite{TH}.  We know that these spinors are orthogonal to each other being  labeled by the angular quantum numbers $ m_{j}$ and
\begin{equation}\label{kjl}
\kappa=\left\{\begin{array}{lcc}
~~~~\,j+\frac{1}{2}=l&{\rm for}& j=l-\frac{1}{2}\\
-(j+\frac{1}{2})=-l-1&{\rm for}& j=l+\frac{1}{2}
\end{array}\right.
\end{equation}
which encapsulates the information about the quantum numbers $l$ and
$j=l\pm\frac{1}{2}$ \cite{TH, LL}.    We note that the antiparticle-like energy eigenspinors  can be obtained directly using the charge conjugation  as in the flat case \cite{C3}.

Thus the spherical variables are separated as  in special relativity  \cite{TH} remaining with a pair of  radial  functions, $f^{\pm}$, (denoted from now without indices) that satisfy a pair of radial equations \cite{CA} that  cannot be solved analytically as it stays. Therefore,  we were forced  to resort to a method of approximation  \cite{CA,CCS} by using a convenient Novikov  dimensionless coordinate \cite{Nov,GRAV}
\begin{equation}\label{x}
x=\sqrt{\frac{r}{r_{0}}-1}\,\in\,(0,\infty)\,,
\end{equation}
and introducing the convenient notations
\begin{equation}
\mu=r_{0}m\,,\quad \varepsilon=r_{0}E\,.
\end{equation}
For very large values of $x$, we can use the Taylor expansion with respect to $\frac{1}{x}$ of the radial Hamiltonian operator,  neglecting  the terms of the order $O(\frac{1}{x^2})$. Thus for $E>0$ we obtain thus  the {\em approximative} scattering radial  solutions  \cite{CA,CCS}
\begin{eqnarray}
\hat f^+(x)&=&C_1^+\frac{1}{x}M_{\rho_+,s}(2i\nu x^2)
+C_2^+\frac{1}{x}W_{\rho_+,s}(2i\nu x^2)\,,\label{E11}\\
\hat f^-(x)&=&C_1^-\frac{1}{x}M_{\rho_-,s}(2i\nu x^2)
+C_2^-\frac{1}{x}W_{\rho_-,s}(2i\nu x^2)\,,\label{E22}
\end{eqnarray}
expressed in terms of Whittaker functions depending on the new parameters  $\nu=\sqrt{\varepsilon^2-\mu^2}$ and
\begin{equation}
s=\sqrt{\kappa^2+\frac{\mu^2}{4}-\epsilon^2},\quad
\rho_{\pm}=\mp\frac{1}{2}-i q,\quad q=\nu+ \frac{\mu^2}{2\nu}.
\end{equation}
The  integration constants must satisfy \cite{CA}
\begin{equation}\label{C1C1}
\frac{C_1^-}{C_1^+}=\frac{s-i q}{\kappa-i\lambda}\,,\quad
\frac{C_2^-}{C_2^+}=-\frac{1}{\kappa-i\lambda}\,,\quad \lambda=\frac{\epsilon\mu}{2\nu}\,.
\end{equation}
These solutions will help us to find the scattering amplitudes of the Dirac particles scattered from different massive bodies after fixing the suitable integration constants.

\section{Partial wave analysis}

In general, a spherically symmetric scattering is described by an energy eigenspinor $U$ whose asymptotic form,
\begin{equation}
U\to U_{plane}(\vec{p})+A(\vec{p},\vec{n}) U_{sph}\,,
\end{equation}
for $r\to \infty$  (where the spacetime becomes flat)  is given by the plane wave spinor of momentum $\vec{p}$ and the free spherical spinors of the flat case  as
\begin{equation}\label{Up}
U_{sph}\propto \frac{1}{r}\,e^{ipr-iEt}\,,\quad p=\sqrt{E^2-m^2}=\frac{\nu}{r_0}\,.
\end{equation}
Here we take $\vec{p}=p\vec{e_3}$ bearing in mind that  in the asymptotic zone the fermion energy is just that of special relativity,$E=\sqrt{m^2+p^2}$. Then  the scattering
angles $\theta$ and $\phi$  are just the spheric angles of the
unit vector $\vec{n}$,  the scattering amplitude
\begin{equation}\label{Ampl}
A(\vec{p},\vec{n})=f(\theta)+ig(\theta)\frac{ \vec{p}\land \vec{n}}{|\vec{p}\land \vec{n}|}\cdot\vec{\sigma}
\end{equation}
depending on  two scalar amplitudes, $f(\theta)$ and $g(\theta)$, that can be  studied by using the partial wave analysis.

For starting our study we consider the asymptotic forms  of our  solutions  (\ref{E11}) and (\ref{E22}) in which the integration constants play the role of free parameters. It is convenient to  separate the general normalization constant $N_{\kappa}$  in Eqs. (\ref{C1C1}) introducing the new constant $C_{\kappa}\in \mathbb{C}$ such that
\begin{eqnarray}
C_1^+&=&N_{\kappa}\,,\qquad \qquad~~~~ C_2^+ = N_{\kappa}C_{\kappa} \,,\\
C_1^-&=&\frac{s-iq}{\kappa-i\lambda}\,N_{\kappa}\,,\qquad C_2^-=-\frac{N_{\kappa}C_{\kappa}}{\kappa-i\lambda}\,.
\end{eqnarray}
The normalization constant is, in general, a complex number that will be denoted from now as
\begin{equation}\label{Nk}
N_{\kappa}=|N_{\kappa}| e^{i\, \frac{\alpha_{\kappa}}{2}}
\end{equation}
where the arbitrary phases $\alpha_{\kappa}$ remain free parameters. Furthermore, taking into account that  $\nu x^2=p(r-r_0)$, we obtain  the asymptotic behavior  of the radial functions of the scattered fermions  as,
\begin{eqnarray}
\left(\begin{array}{l}
f^+\\
f^-
\end{array}\right)&=&\left(
\begin{array}{c}i\sqrt{\varepsilon+\mu}\,(\hat f^- -\hat f^+)\\
\sqrt{\varepsilon-\mu}\,(\hat f^- +\hat f^+)
\end{array}\right)\nonumber\\
&\propto& \begin{array}{c}
\sqrt{E+m}\,\sin\\
\sqrt{E-m}\,\cos
\end{array}\left( pr-\frac{\pi l}{2} +\delta_{\kappa}+\vartheta(r)\right)\,.\label{solfin}
\end{eqnarray}
The point-independent phase shifts $\delta_{\kappa}$  can be derived comparing this asymptotic behavior with the ratio $\frac{\hat f_a^+}{\hat f_a^-}$ of the asymptotic radial  functions (\ref{As1}) and (\ref{As2}) given in the Appendix A. Thus we obtain the closed form \cite{CCS1}
\begin{equation}\label{final}
S_{\kappa}=e^{2i\delta_{\kappa}}=\frac{\kappa-i\lambda}{s+iq}\,\frac{\frac{\Gamma(1+2s)}{\Gamma(s+iq)}
e^{i\pi(l-s)}}{\frac{\Gamma(1+2s)}{\Gamma(s-iq)}-C_{\kappa} e^{-i\pi(s+iq)}}\,,
\end{equation}
depending on the parameter $C_{\kappa}$ which can take different values in each partial wave. Notice that the values of  $\kappa$ and $l$ are related as in Eq. (\ref{kjl}), i. e.
\begin{equation}\label{l}
l=|\kappa|-\frac{1}{2}(1-{\rm sign}\,\kappa)
\end{equation}
The remaining point-dependent phase,
\begin{equation}
\vartheta(r)= -p r_0+q \ln [2p(r-r_0)]\,,
\end{equation}
does not depend on angular quantum numbers and therefore it may be ignored as
in the  Dirac-Coulomb case \cite{LL,S3}.

The final result (\ref{final}) depending on the parameters  introduced above that can be expressed in terms of physical quantities by using the second of Eqs. (\ref{Up}) as,
\begin{eqnarray}
s&=&\sqrt{\kappa^2-k^2}\,,\quad k=M\sqrt{4p^2+3m^2}\,,\label{s}\\
q&=&\frac{M}{p}\,(2p^2+m^2)\,,\label{q}\\
\lambda&=&\frac{M}{p} m\sqrt{m^2+p^2}\,. \label{lam}
\end{eqnarray}
The parameters $k,q,\lambda \in {\Bbb R}^+$ are  positively defined
and satisfy the identity
\begin{equation}\label{kqlam}
k^2=q^2-\lambda^2\,,
\end{equation}
while the parameter $s$ can take either real  values  or pure imaginary ones.
In the case of the massless fermions ($m=0$) we remain
with  the unique parameter $k=q=2pM$ since then $\lambda=0$.

In general, the phase shifts $\delta_{\kappa}$ may be real or complex numbers such that
$|S_{\kappa}| \le 1$. When $|S_{\kappa}|=1$ the scattering is {\em elastic} while for $|S_
{\kappa}|<1$ a part of the incident beam is absorbed.  Here we study only the elastic scattering with real valued phase shifts for which the scalar amplitudes of Eq. (\ref{Ampl}),
\begin{eqnarray}
f(\theta)=\sum_{l=0}^{\infty}a_l\,P_l(\cos \theta)\,,\label{f} \qquad  g(\theta)=\sum_{l=1}^{\infty}b_l\,P_l^1(\cos\theta)\,, \label{g}
\end{eqnarray}
depend on the following {\em partial} amplitudes \cite{LL,S3},
\begin{eqnarray}
a_l&=&(2l+1)f_l\nonumber=\frac{1}{2ip}\left[(l+1)(S_{-l-1}-1)+l(S_l-1)\right]\,,\label{fl}\\
b_l&=&\!\!\!(2l+1)g_l=\frac{1}{2ip}\left(S_{-l-1}-S_l\right)\,.\label{gl}
\end{eqnarray}
These amplitudes give rise to  the {elastic} scattering intensity or  differential cross section,
\begin{equation}\label{int}
{\cal I}(\theta)=\frac{d\sigma}{d\Omega}=|f(\theta)|^2+|g(\theta)|^2\,,
\end{equation}
and the polarization degree of the scattered fermions,
\begin{equation}\label{pol}
{\cal P}(\theta)=-i\frac{f(\theta)^*g(\theta)-f(\theta) g(\theta)^*}{|f(\theta)|^2+|g(\theta)|^2}.
\end{equation}
This last quantity is interesting for the scattering of massive fermions  representing the induced polarization of the scattered fermions for an unpolarized incident beam.

Note  that all the quantities derived here depend on the fixed $E$ (or $p$). When we intend to investigate a domain of energies then we have to speak about functions of $E$ (or $p$)  as for example $S_{\kappa}(E), C_{\kappa}(E),...$ etc.

\section{Boundary conditions and phase shifts}

In what follows we consider that the target is a massive spherical body of mass $M$ and radius $R\gg r_0$. In these circumstances we may study the scattering of the Dirac fermions on this target by using exclusively our asymptotic solutions  with different boundary conditions on the surface of the radius $R$ delimiting the body. These boundary conditions may depend on the properties of the exterior surface which may reflect, absorb or  polarize the scattered beam. In our formalism, these properties must be encapsulated in the form of the boundary conditions determining the constant $C_\kappa$ of the spinor solutions and, implicitly, the scattering amplitudes.

In Refs. \cite{CCS} and \cite{CCS1} we have shown that in the case of the genuine Schwarzschild and Reissner-Nordstr\" om black holes we must take $C_\kappa=0$. Then for the Schwarzschild black holes we obtain an elastic scattering for $|\kappa| > {\rm floor}(k)$ and  absorption for  $1\le |\kappa|\le {\rm floor}(k)$. A similar result holds for the Reissner-Nordstr\" om black holes.

However, in the case of of a massive body of radius $R>r_0$ we must look for  more sophisticated boundary conditions according to the particular  physical properties of the spherical body. The only tool we have in our investigation is the form of the partial radial currents \cite{CCS}
\begin{eqnarray}
J_{rad}&=&i\left( f^+ {f^-}^*-{f^+}^*f^-\right)\nonumber\\
&=&2 p\,r_0 \left(|\hat f^+|^2-|\hat f^-|^2\right)
= -p\,r_0\, \sinh\left(2\Im \delta_{\kappa}\right)\,, \label{Jrad}
\end{eqnarray}
produced by the particular solutions (\ref{solfin}) where $\nu=p\,r_0$. These are conserved quantities which do not depend explicitly on $r$  since the condition $\partial_r J_{rad}=0$ is fulfilled whenever the functions $f^{\pm}$ are solutions of the radial equations \cite{CCS}.

The most general boundary condition may be a general linear relation between these two functions of the form
\begin{equation}
\left.\hat f^-(r)^*\right|_{r=R} = a\left.\hat f^+(r)\right|_{r=R}+b\left. \hat f^-(r)\right|_{r=R}\,,
\end{equation}
where $a$ and $b$ are complex numbers.  The condition of preventing absorption is very selective leaving us with two possibilities
\begin{equation}
J_{rad}=0 ~~ \to ~~\left\{
\begin{array}{ll}
a=e^{i\phi_a}\,,&~b=0\\
a=0\,,&~b=e^{i\phi_b}
\end{array}\right.
\end{equation}
On the bother hand, in the case of the elastic scattering the functions $f^+ $ and $f^-$  must have the same phase, which forces us to chose $b=0$ and $a=1$  remaining with the simple boundary condition
\begin{equation}\label{bc}
\left.\hat f_{E,\kappa}^-(r)^*\right|_{r=R} = \left.\hat f_{E,\kappa}^+(r)\right|_{r=R}
\end{equation}
in each partial wave where we intend to prevent absorption. Then  the constants $C_{\kappa}(E)$ can be obtained by solving the Eq. (\ref{bc}) where the functions $\hat f^+$ and $\hat f^-$ for $r=R$ are given by Eqs. (\ref{E11}) and (\ref{E22}) in which we must take
\begin{equation}
x=x_R=\sqrt{\frac{R}{r_0}-1}\,.
\end{equation}
Thus our boundary conditions determine completely the constants $C_{\kappa}(E)$ for all the values of  $\kappa$ of the partial waves in which we desire to prevent absorption.

In what follows we focus on the simplest case of the total reflection when the scattering is elastic in any partial waves. Therefore, we assume that the boundary condition (\ref{bc}) is satisfied in any partial wave while the parameter $\alpha$ is independent on $\kappa$ being determined only by the properties of the exterior surface of the massive body. Under such circumstances we do not need to resort to numerical investigation since the problem can be solved analytically by using the asymptotic radial functions (\ref{As1}) and (\ref{As2})  which helped us to obtain the result (\ref{final}).  The main task is to derive the constant $C_{\kappa}$ from the condition (\ref{bc}) rewritten in terms of these functions  as
\begin{equation}
\left.\hat f_a^-(x)^*\right|_{x=x_R}=\left.\hat f_a^+(x)\right|_{x=x_R}\,.
\end{equation}
In this equation the parameter $s$ can take either real values when $|\kappa|>{\rm floor} (k)$ or pure imaginary ones $s=\pm i|s|$ if $1\le |\kappa|\le {\rm floor} (k)$. These three cases must be considered separately denoting from now the quantities  $C_{\kappa}$ and the corresponding $S_{\kappa}$ given by Eq. (\ref{final})  as
\begin{eqnarray}
C_{\kappa}^>~\to~S_{\kappa}^>\,, &~ {\rm for}~& s=|s| ~~~~\in {\Bbb R}\,,\\
C_{\kappa}^{\pm}~\to~S_{\kappa}^{\pm}\,, &~ {\rm for}~& s=\pm i |s| ~~\in {\Bbb C}\,.
\end{eqnarray}
From the asymptotic form of Eq. (\ref{bc}) we obtain first
\begin{eqnarray}\label{Cg}
C_{\kappa}^>&=&e^{-\pi q}\frac{\Gamma(1+2s)}{\Gamma(1+s-iq)}\left[(\lambda+i\kappa)e^{-i\alpha_{\kappa}}-(s-iq)e^{i\pi s}\right]\,,
\end{eqnarray}
giving the simple closed form
\begin{equation}\label{Smare}
S_{\kappa}^>=i \,\frac{\Gamma(1+s-iq)}{\Gamma(1+s+iq)}\,e^{i(\pi l+\alpha_{\kappa})}\,, \quad \forall s\in {\Bbb R}\,,
\end{equation}
For the imaginary values  $s=\pm i|s|$ we derive two different constants
\begin{eqnarray}
C_{\kappa}^{\pm}&=&-i\,\frac{\Gamma(1\pm 2i|s|)}{\Gamma(1\pm i|s|-iq)}(q\pm|s|)e^{-\pi (q\pm |s|)}\nonumber\\
&&+\frac{\Gamma(1\mp 2i|s|)}{\Gamma(1\mp i|s|-iq)}(\lambda+i\kappa)e^{-\pi q-i\alpha_{\kappa}}\,,
\end{eqnarray}
which satisfy $C_{\kappa}^{-}(|s|)=C_{\kappa}^{+}(-|s|)$ and, fortunately, lead to the same unique result,
\begin{eqnarray}\label{Smic}
S_{\kappa}^<&=&S_{\kappa}^+=S_{\kappa}^-\nonumber\\
&=&i \,\frac{\Gamma(1-i|s|-iq)}{\Gamma(1+i |s|+iq)}\frac{\Gamma(1+2i|s|)}{\Gamma(1-2 i |s|)}\,e^{i(\pi l+\alpha_{\kappa})}\,.
\end{eqnarray}
Summarizing, we may write the final result as
\begin{equation}
S_{\kappa}=e^{2i\delta_{\kappa}}=\left\{
\begin{array}{llr}
S_{\kappa}^<&~~{\rm for}&~~1\le |\kappa|\le {\rm floor} (k)\\
S_{\kappa}^>&~~{\rm for}& |\kappa|> {\rm floor} (k)
\end{array} \right.
\end{equation}
where $S_{\kappa}^<$ and $S_{\kappa}^>$ are given by Eqs. (\ref{Smic}) and respectively (\ref{Smare}).

Thus we succeeded to solve  the problem of the total reflection on a massive body in the asymptotic approximation. It is obvious that our boundary condition prevents absorption since $|S_{\kappa}|=1$ for all the possible values of $\kappa=\pm1,\pm2,...$. Moreover we observe that $S_{\kappa}$ behaves as a continuous function of $s$ since in the branch point $s=0$ we have
\begin{equation}
\lim_{s\to 0}S_{\kappa}^<=\lim_{s\to 0}S_{\kappa}^>=i \,\frac{\Gamma(1-iq)}{\Gamma(1+iq)}\,e^{i(\pi l+\alpha_{\kappa})}\,.
\end{equation}
Another obvious property is $S_{-\kappa}=S_{\kappa}$ since these quantities depend only on $k^2$ only as functions of  $s$.

It is important to note that the above results are in accordance with our previous study of the fermions scattered from a bare black hole \cite{CCS} for which we used the general boundary condition $C_{\kappa}=0$ for any $s$. Indeed, for $s=|s|$  the constant  (\ref{Cg}) can be canceled as,
\begin{equation}\label{naked}
C_{\kappa}=0~~\to~~e^{i\alpha_{\kappa}}=\frac{\lambda- i \kappa}{s-iq}e^{-i\pi s} \,,
\end{equation}
as long as we have $|\lambda- i \kappa|=|s-iq|$ as it results from Eqs. (\ref{s}-\ref{kqlam}). Thus we recover the result of Ref. \cite{CCS} for $s\in {\Bbb R}$ while for imaginary values of $s$ the boundary condition (\ref{bc}) is no longer satisfied allowing absorption.

Finally, we must specify that  in the asymptotic approximation used here we neglect the terms of the order ${\cal O}(\frac{1}{x^2})$ of the radial functions and implicitly the terms  of the order ${\cal O}(\frac{1}{R})$ in Eq. (\ref{bc}) such that the final result is independent  on $R$. This corresponds  to the ideal image of the target which is seen as  a point-like massive particle.  This image can be corrected by resorting to additional numerical methods but, as mentioned, these corrections are very small and less relevant.

\section{Graphical analysis and discussion}

Let us now discuss some physical consequences of our results concerning the fermion elastic scattering  from a massive body which reflects totally the incident beam. We use the phase shifts given by Eqs.  (\ref{Smare}) and (\ref{Smic})  where the phases $\alpha_{\kappa}$ cannot be determined in the ideal approximation considered here. Therefore, these remain free parameters related to the properties of the reflecting surface of the massive body.  Bearing in mind that the absence of this surface leads to the phases (\ref{naked}) we understand that we have the opportunity of building different models, postulating the form of the phases $\alpha_{\kappa}$ and analyzing then {\em a posteriori} the physical effects supposed to be due to the reflecting surface.

\begin{figure}
\centering
\includegraphics[scale=0.45]{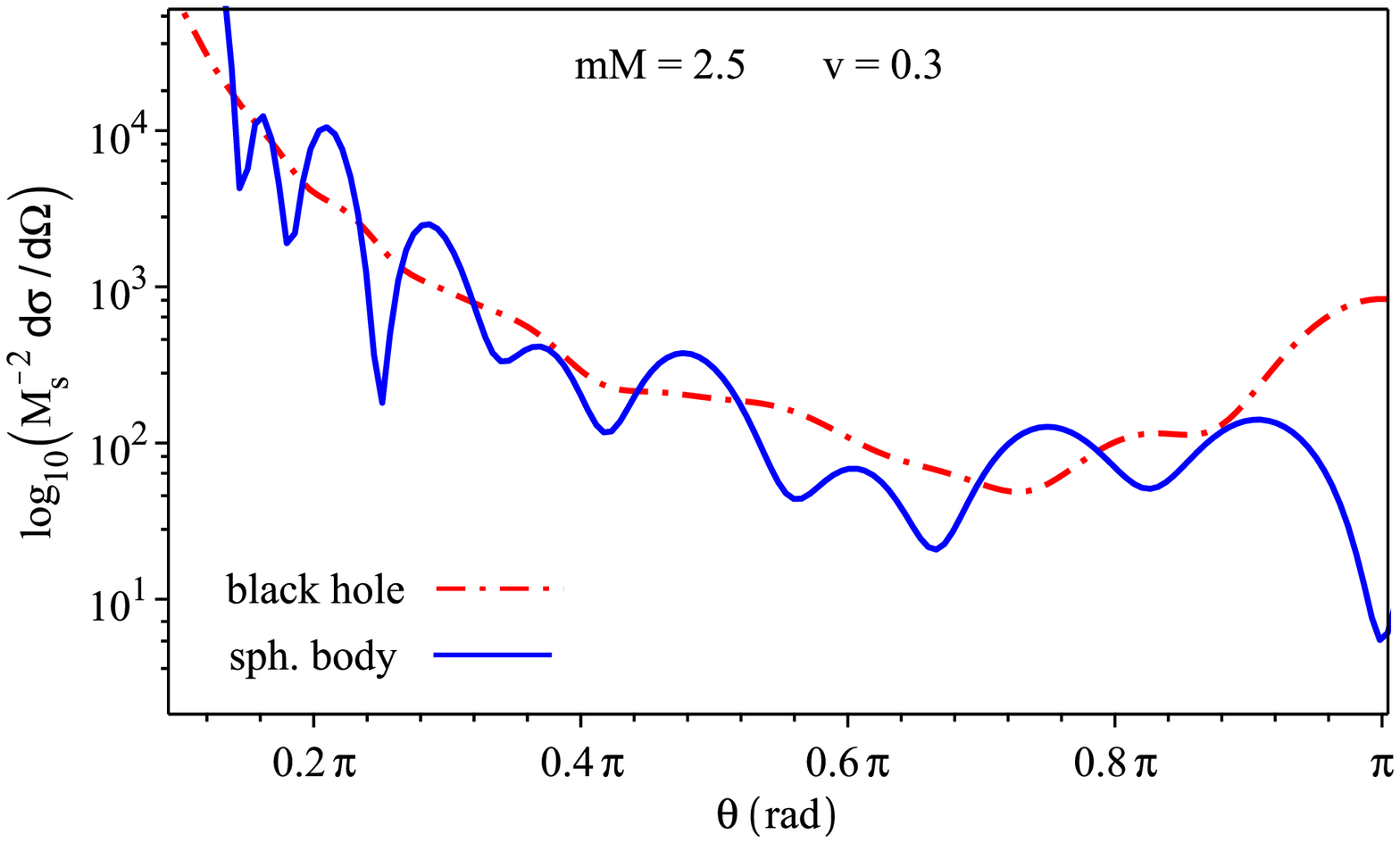}
\includegraphics[scale=0.45]{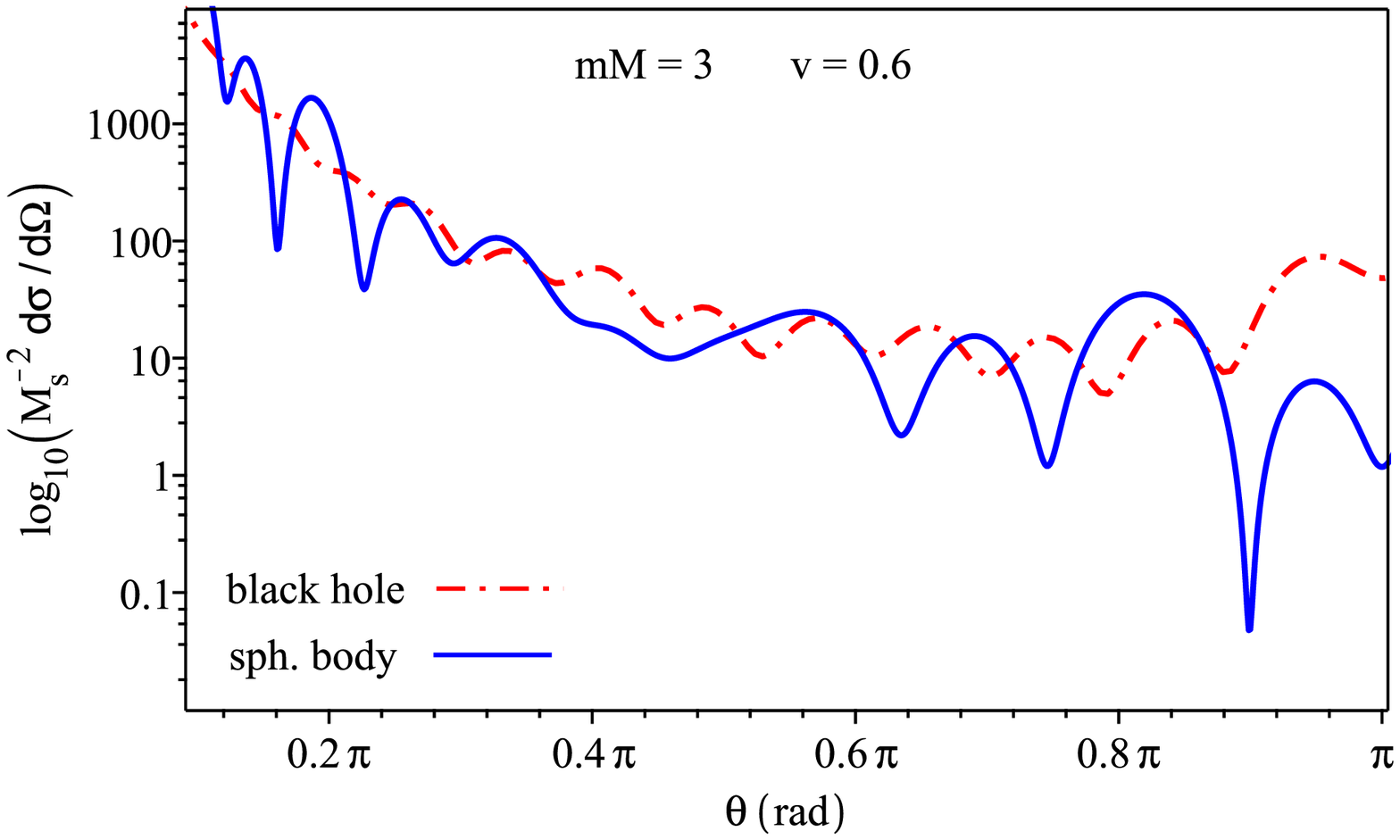}
\includegraphics[scale=0.45]{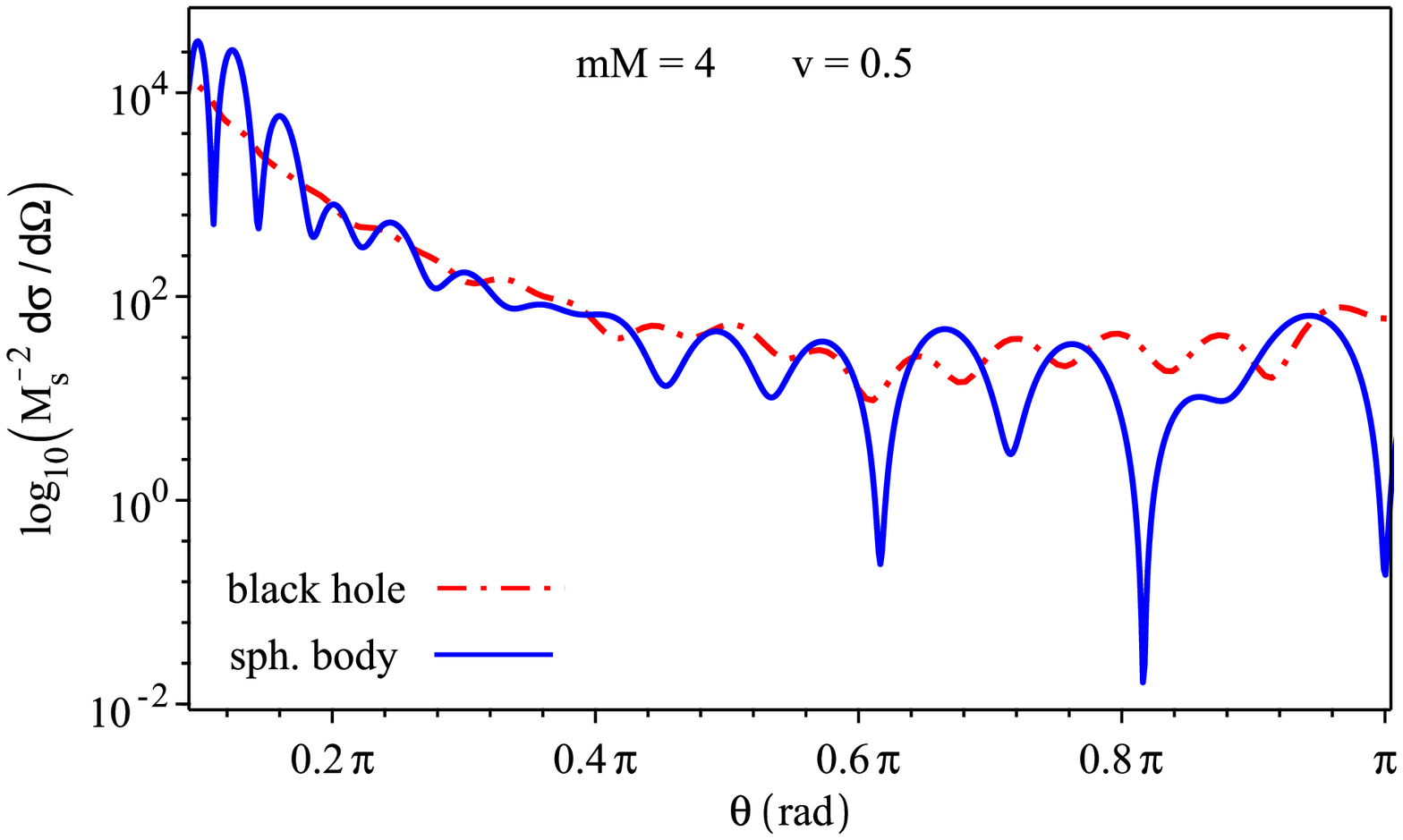}
\caption{ Comparison between the scattering by a massive spherical symmetric body (blue curves) and scattering by a Schwarzschild black hole (dash-dotted curves). In the backward direction ($\theta\approx\pi$) one can observe the presence of a glory. The oscillations in the scattering intensity indicates spiral (orbiting) scattering, which is much more significant for a massive spherical body than for a black hole.}
\label{fig1}
\end{figure}

Obviously, this analysis is laborious and cannot be accomplished here in few pages such that  we restrict ourselves to discuss  a simple  simple model in which the phase shifts depend on $\kappa$ only through $s$. Therefore, in this model we must take
\begin{equation}\label{aa}
\alpha_{\kappa}=\pm \pi l
\end{equation}
where $l$ is given by Eq. (\ref{l}) for any given $\kappa$.  In what follows we analyze this model focusing on the scattering intensity and induced polarization as functions of the  scattering angle $\theta$.

In general the amplitudes of the method of partial wave are singular for $\theta=0$ such that we must apply the method proposed by Yennie et all \cite{Yeni} for removing them. Therefore, we proceed as  in Ref.\cite{S3,CCS,CCS1} by replacing the series  (\ref{f}) with the
reduced ones of $m$-th order defined as,
\begin{eqnarray}
f(\theta)&=&\frac{1}{(1-\cos\theta)^{m_1}} \sum_{l\ge 0} a_l^{(m_1)} P_l(cos\theta)\,,\label{gf}\\
g(\theta)&=&\frac{1}{(1-\cos\theta)^{m_2}} \sum_{l\ge 1} b_l^{(m_2)} P_l^1(\cos\theta)\,.\label{gf1}
\end{eqnarray}
The recurrence relations satisfied by the Legendre polynomials
$P_{l}(x)\,,P_{l}^{1}(x)$ lead  to  the iterative rules giving the reduced coefficients in any order
\begin{eqnarray}
a_l^{(i+1)}&=&a_l^{(i)}-\frac{l+1}{2l+3}a_{l+1}^{(i)}-\frac{l}{2l-1}a_{l-1}^{(i)}\,,\\
b_l^{(i+1)}&=&b_l^{(i)}-\frac{l+2}{2l+3}b_{l+1}^{(i)}-\frac{l-1}{2l-1}b_{l-1}^{(i)}\,,
\end{eqnarray}
if we start with $a_l^{(0)}=a_l$ and $b_l^{(0)}=b_l$ as defined by Eqs. (\ref{fl}). We use $m_1=2$ and $m_2=1$ for obtaining the plots.

\begin{figure}
\centering
\includegraphics[scale=0.45]{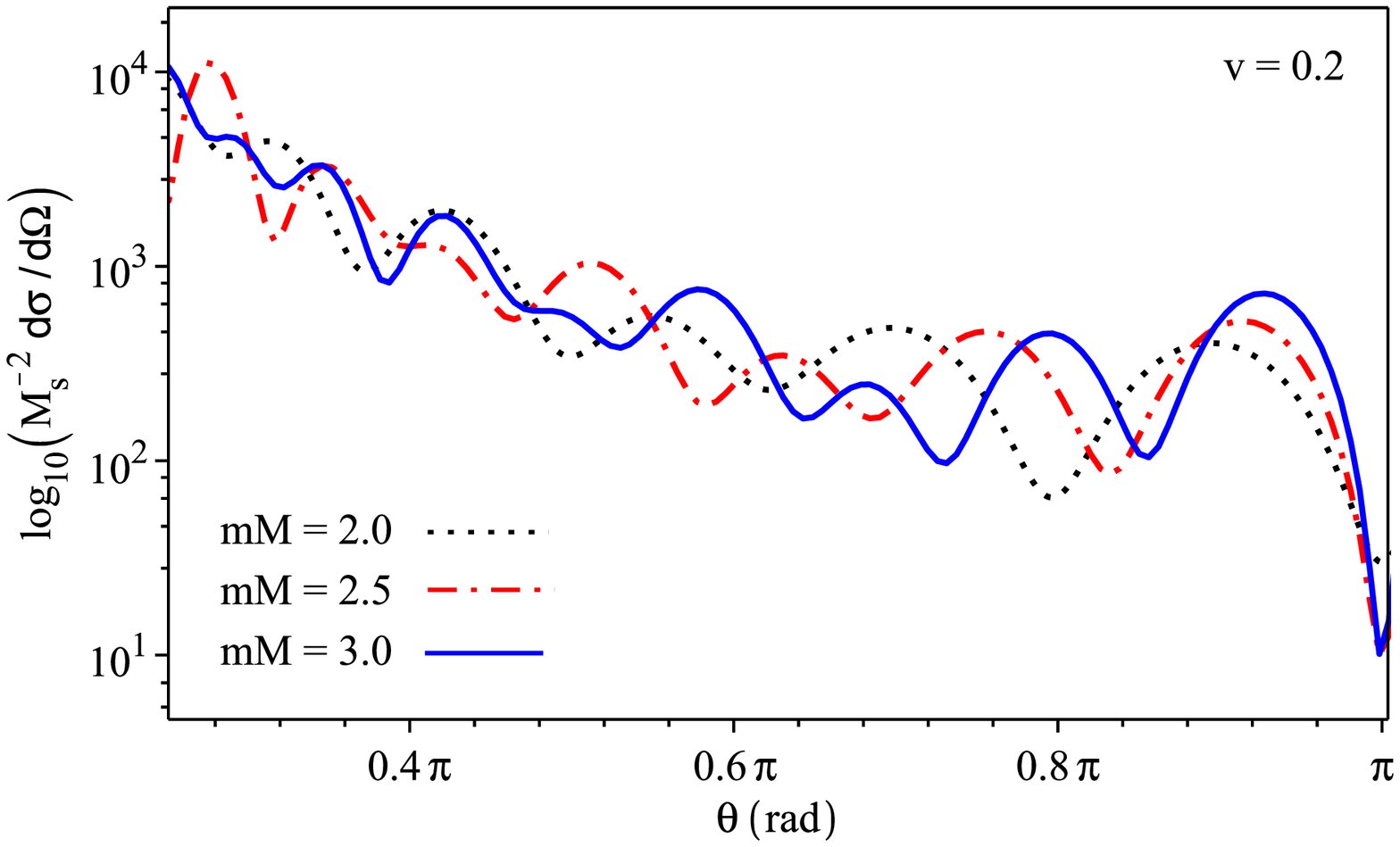}
\includegraphics[scale=0.45]{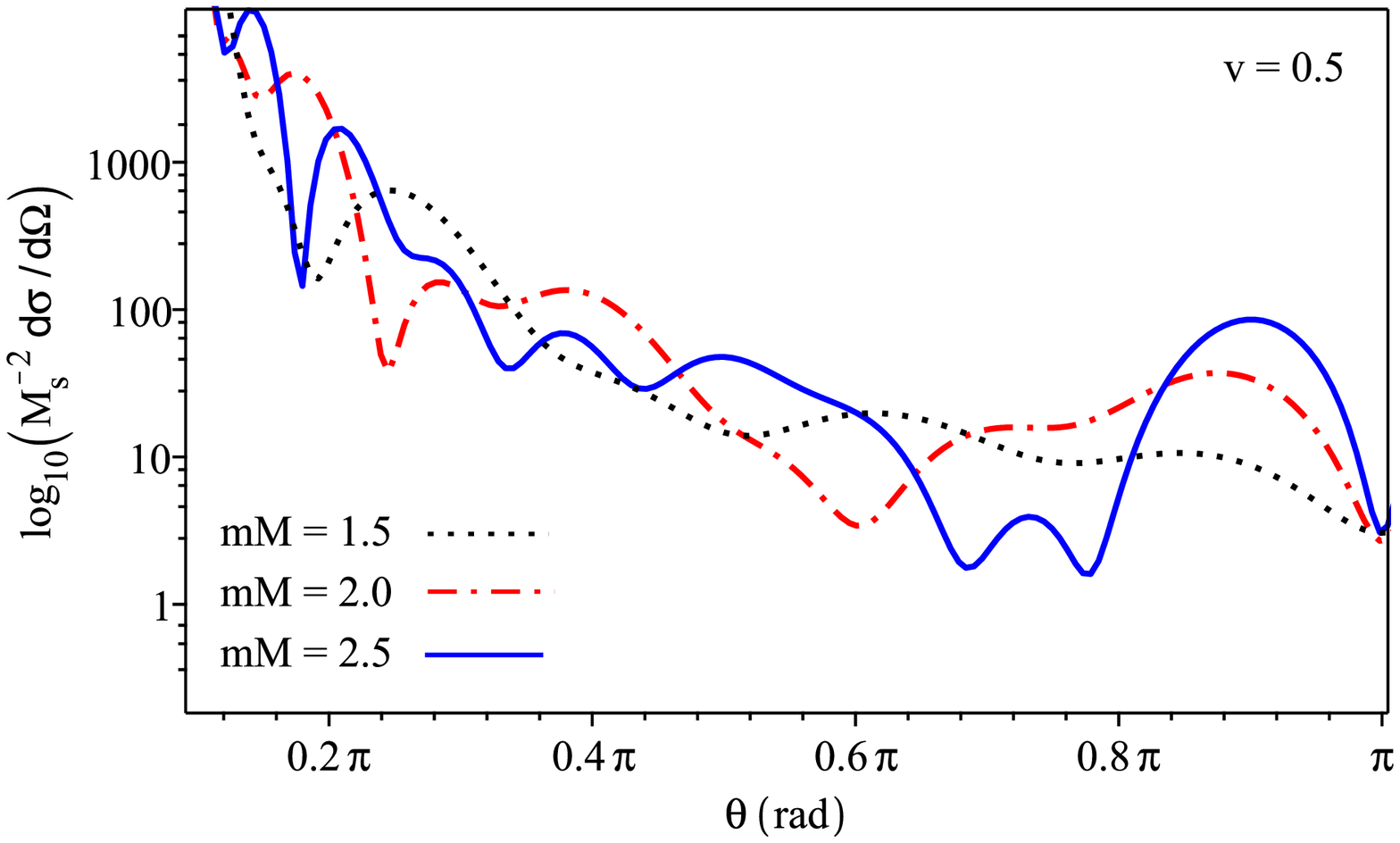}
\includegraphics[scale=0.45]{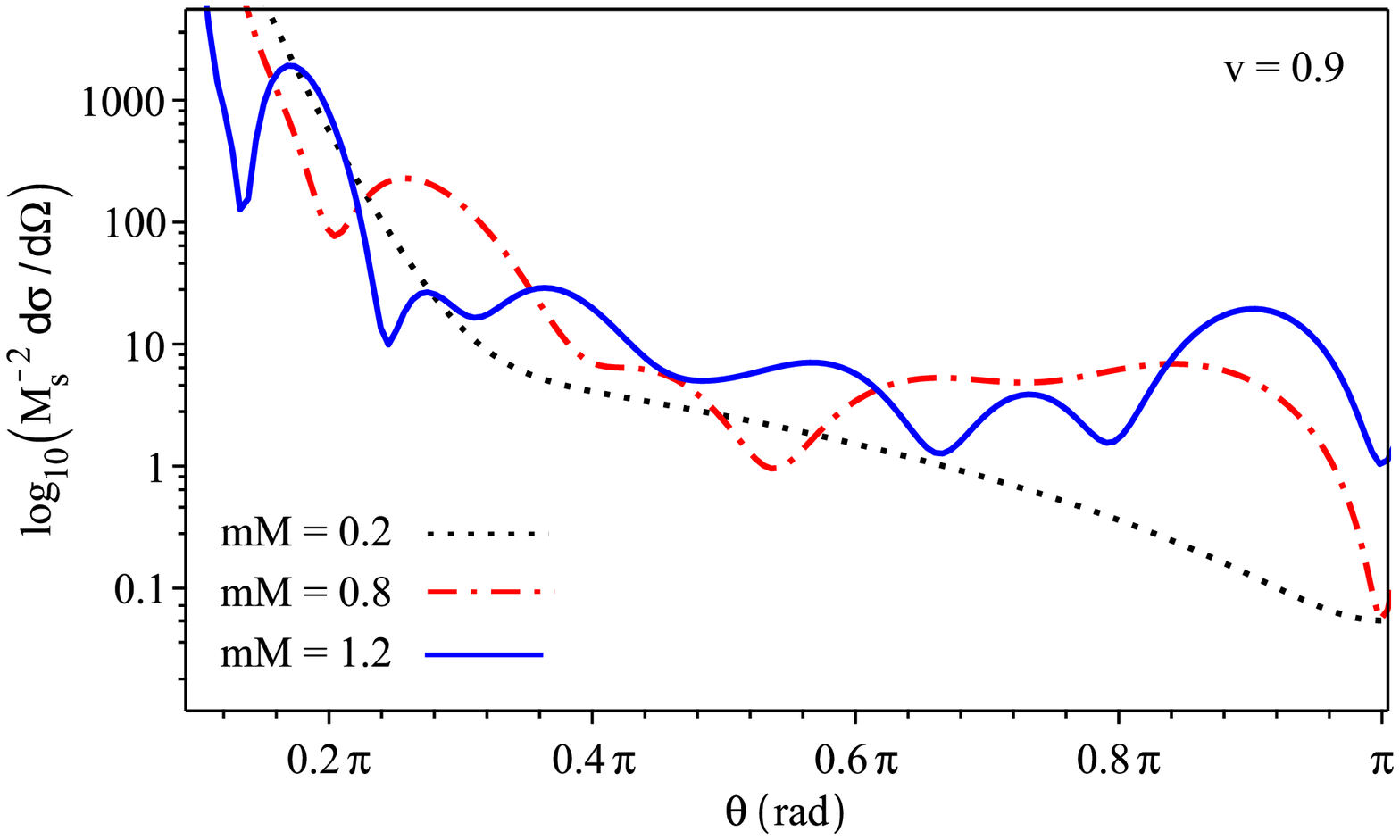}
\caption{Glory and spiral scattering by a massive spherical body over a range of gravitational couplings $mM$ and a fixed value of fermion's speed $v$. The glory magnitude and angular frequency of spiral scattering are enhanced with the body's mass $M$.}
\label{fig2}
\end{figure}

In Fig. \ref{fig1} we compare the scattering from a Schwarzschild black hole (red dash-dotted curves) with that of a spherical symmetric massive body (blue continuous curves). The first ting to observe is that the scattering by a body has always a minima on-axis in the backward direction ($\theta\approx\pi$) regardless of the fermion's speed in contrast with scattering by a black hole that has a minima on-axis in the case of relativistic fermions and respectively a maxima on-axis for nonrelativistic fermions. This implies the presence of a halo in the glory when scattering by a body and a bright spot or hallo in the glory of a black hole. Furthermore, the magnitude of oscillations in the scattering intensity is much higher for scattering by a body than by a black hole. The direct consequence of this is the fact that the massive spherical body cross section has much more pronounced spiral scattering, or orbiting, (i.e. the presence of oscillations at intermediate angles in the scattering cross section). Analysing the behaviour of the scattering cross section at small scattering angles we observe that spiral scattering is present for the case of massive spherical body, whereas for the black hole case is missing or it is very small (see also Fig. \ref{fig2}).

Fig. \ref{fig2} presents plots of the scattering cross section for a fixed value of the fermion's speed $v$ (in units of $c$) and three different values of the parameter $mM$, that can be related to the quantity $\epsilon=\frac{GME}{\hbar c^3}=\frac{\pi r_S}{v\lambda}$ which gives a convenient dimensionless measure of the gravitational coupling. The relation is $mM=ME\sqrt{1-v^2}$ and also $r_S$ stands for the Schwarzschild radius, while $\lambda=h/p$ is the quantum wavelength.

Analyzing Fig. \ref{fig2} one can observe that the magnitude and the angular frequency of spiral scattering is increasing with the mass of the spherical body. Furthermore, the glory is also enhanced with increasing $M$. We also notice that the scattering intensity takes higher values for non-relativistic fermions compared with relativistic ones. The spiral (orbiting) scattering is not present in the cross section at low values of $mM$. However, as the value of $mM$ is increased more complex scattering patterns start to appear.  Looking at the glory peek we see that it is increasing with $mM$ and at the same time the width it's narrowing down.

\begin{figure}
\centering
\includegraphics[scale=0.45]{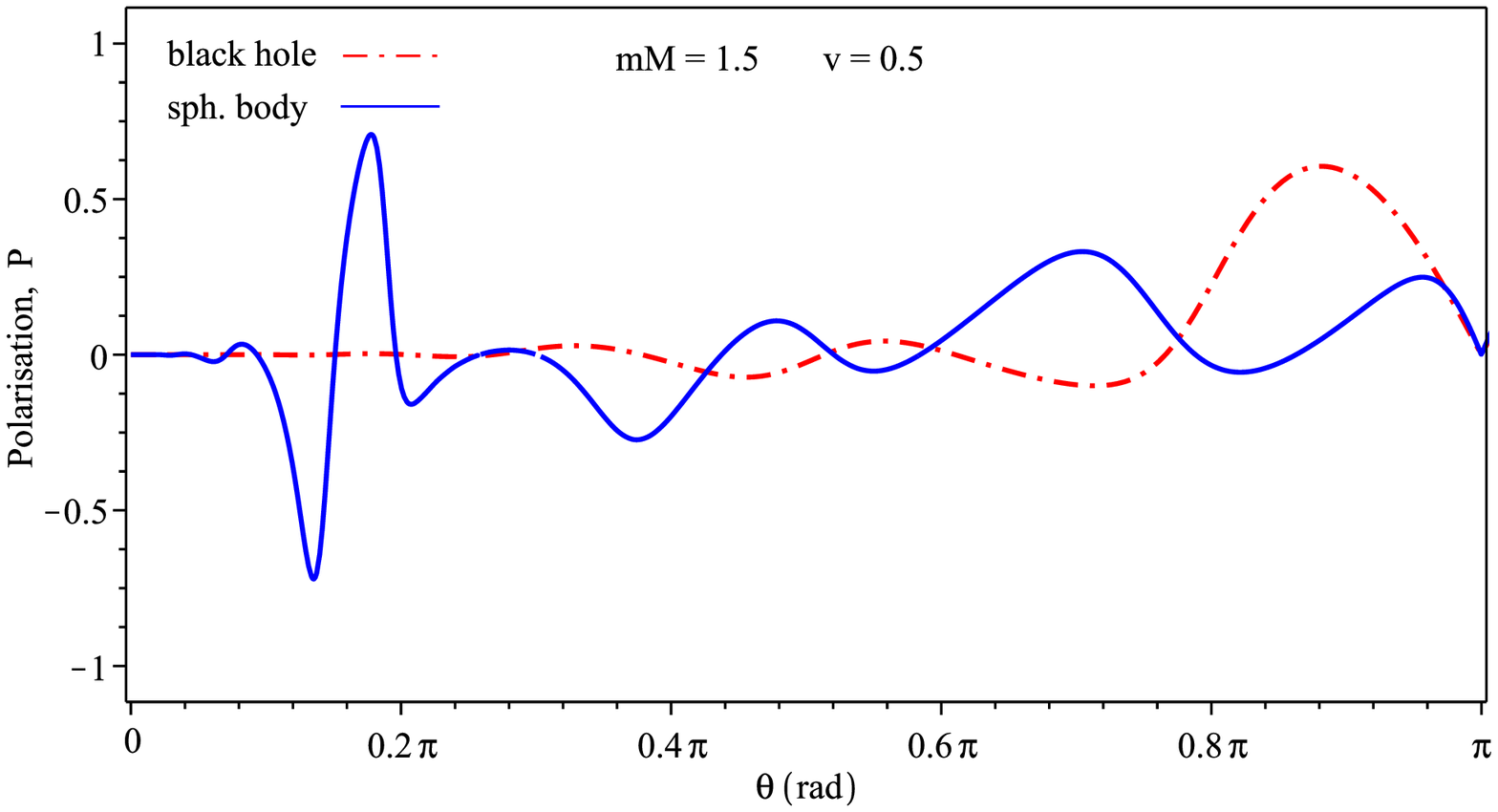}
\includegraphics[scale=0.45]{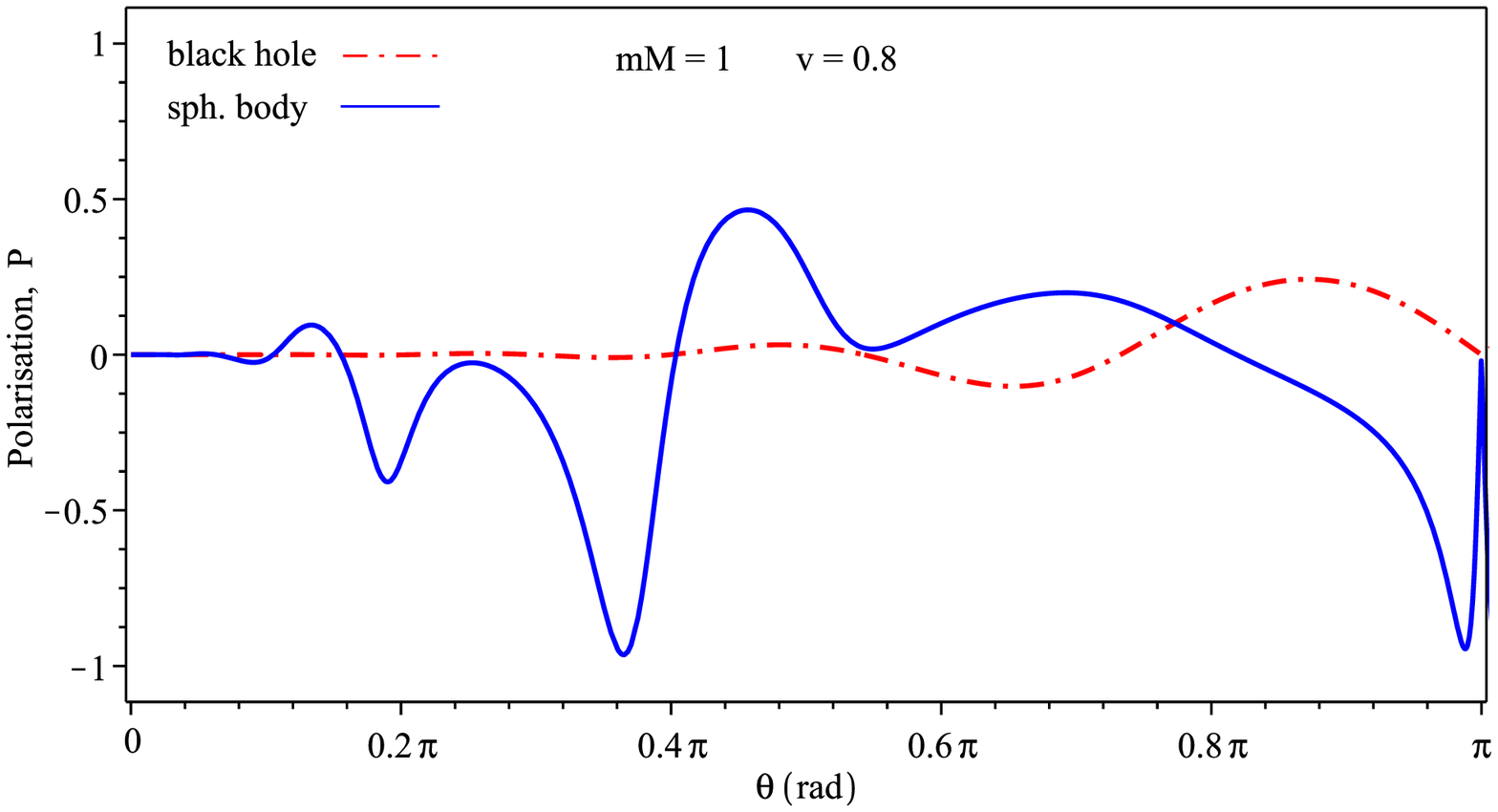}
\includegraphics[scale=0.45]{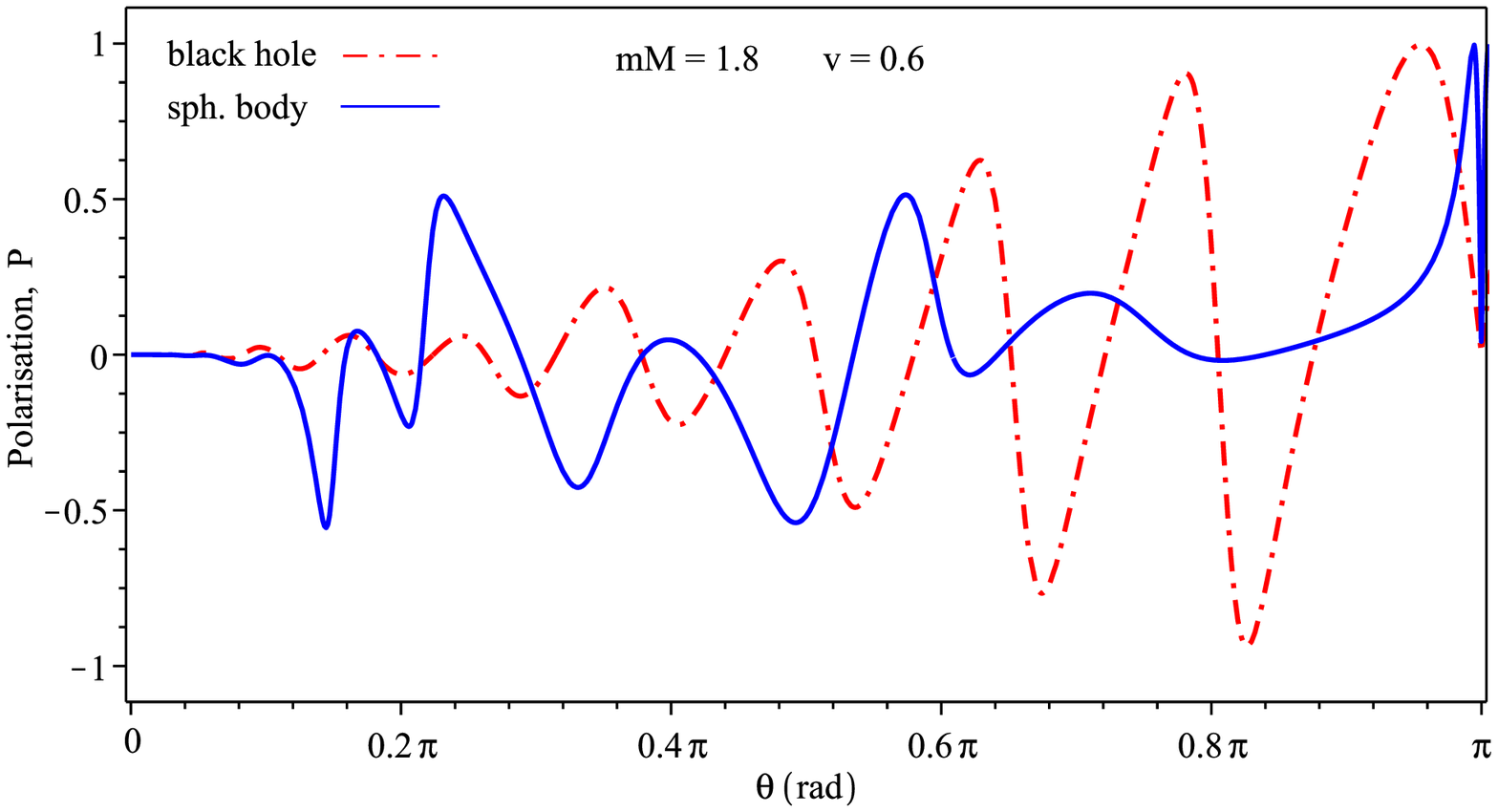}
\caption{ Polarization as a function of the scattering angel $\theta$. Comparatione with the black hole polarization. We observe that the pattern of polarization for the massive spherical body is much more complex (solid blue curves). }
\label{fig3}
\end{figure}

As showed in Refs. \cite{S3,CCS,CCS1}  an initially unpolarized beam of incident fermion beam could become partially polarized after the interaction with a black hole. This conclusion remains valid also for the gravitational interaction with a massive spherical body.  Fig. \ref{fig3} shows how the polarization varies with the scattering angle for a given value of the speed $v$ and of the parameter $mM$.  We observe that the massive spherical body generates more complex patterns in the polarization in comparison with the black hole case. The oscillations that appear in the polarization can be correlated with the oscillations present in the scattering cross section, that give rise to glory and spiral scattering.

\section{Concluding remarks}

We presented here a simple model of massive body surrounded by a surface able to reflect totally the incident beam of massive Dirac fermions. We used an asymptotic approximation  which is suitable for developing the partial wave analysis in terms of simple closed formulas giving the phase shifts that allowed us to study the principal features of the fermion scattering from the massive bodies reflecting the incident beam.  Thus we may compare the scattering from massive bodies with that from black holes finding significant differences in what concerns the profile of the scattering intensity and induced polarization. Thus we may conclude that our analytical method is accurate enough for revealing the principal differences among the massive bodies and bare black holes.

However, our approach can be refined by using numerical methods for improving the boundary conditions on exterior reflecting surfaces where some additional physical effects may be considered.  For this reason our further objective is to complete our analytical approach with effective numerical methods for studying more complicated scattering processes.

\begin{acknowledgements}
C.A. Sporea was supported by a grant of Ministery of Research and Innovation, CNCS - UEFISCDI, project number PN-III-P1-1.1-PD-2016-0842, within PNCDI III.
I. I. Cot\u aescu was partially supported by a grant of  the Romanian Ministry of Research and Innovation, CCCDI-UEFISCDI, project number  PN-III-P1-1.2-PCCDI-2017-0371, within PNCDI III.
\end{acknowledgements}

\appendix

\section{Asymptotic radial functions}

The asymptotic form of the Whittaker functions $W_{\kappa,\mu}(z)$ and
\begin{eqnarray}
M_{\kappa,\mu}(z)&=&\frac{\Gamma(1+2\mu)}{\Gamma(\frac{1}{2}+\mu+\kappa)}e^{-i\pi(\kappa-\mu-\frac{1}{2})}W_{\kappa,\mu}(z)\nonumber\\
&&+\frac{\Gamma(1+2\mu)}{\Gamma(\frac{1}{2}+\mu-\kappa)}e^{-i\pi\kappa}W_{-\kappa,\mu}(-z)\,,
\end{eqnarray}
can be obtained approximating \cite{NIST}
\begin{equation}
W_{\kappa,\mu}(z)\sim e^{-\frac{1}{2}z} z^{\kappa}\,.
\end{equation}
Substituting this approximation of the Whittaker functions in Eqs. (\ref{E11}) and (\ref{E22}) and neglecting the terms of the order ${\cal O}(\frac{1}{x^2})$ we obtain the asymptotic solutions used here
\begin{eqnarray}
\hat f^+_a(x)&=&N_{\kappa}i e^{-\pi q}\frac{\Gamma(1+2s)}{\Gamma(1+s+iq)}\frac{e^{i\nu x^2}(-2i\nu x^2)^{iq+\frac{1}{2}}}{x}\,,\label{As1}\\
\hat f^-_a(x)&=&N_{\kappa}\left[\frac{s-iq}{\kappa-i\lambda}\frac{\Gamma(1+2s)}{\Gamma(1+s-iq)}e^{-\pi q+i\pi s} -\frac{C_{\kappa}}{\kappa-i\lambda}\right]\nonumber\\
&&\times\frac{e^{-i\nu x^2}(2i\nu x^2)^{-iq+\frac{1}{2}}}{x}\,,\label{As2}
\end{eqnarray}
where we know that $\nu x^2=p(r-r_0)$ and $N_{\kappa}$ is the normalization constant (\ref{Nk}).

\end{document}